\documentstyle[preprint,aps,epsf]{revtex}

\def\singlespace {\smallskipamount=3.75pt plus1pt minus1pt
                  \medskipamount=7.5pt plus2pt minus2pt
                  \bigskipamount=15pt plus4pt minus4pt
                  \normalbaselineskip=12pt plus0pt minus0pt
                  \normallineskip=1pt
                  \normallineskiplimit=0pt
                  \jot=3.75pt
                  {\def\smallskip {\vskip\smallskipamount}}
                  {\def\medskip   {\vskip\medskipamount}}
                  {\def\bigskip   {\vskip\bigskipamount}}
                  {\setbox\strutbox=\hbox{\vrule
                    height10.5pt depth4.5pt width 0pt}}
                  \parskip 7.5pt
                  \normalbaselines}
\def\middlespace {\smallskipamount=5.625pt plus1.5pt minus1.5pt
                  \medskipamount=11.25pt plus3pt minus3pt
                  \bigskipamount=22.5pt plus6pt minus6pt
                  \normalbaselineskip=22.5pt plus0pt minus0pt
                  \normallineskip=1pt
                  \normallineskiplimit=0pt
                  \jot=5.625pt
                  {\def\smallskip {\vskip\smallskipamount}}
                  {\def\medskip   {\vskip\medskipamount}}
                  {\def\bigskip   {\vskip\bigskipamount}}
                  {\setbox\strutbox=\hbox{\vrule
                    height15.75pt depth6.75pt width 0pt}}
                  \parskip 11.25pt
                  \normalbaselines}
\def\doublespace {\smallskipamount=7.5pt plus2pt minus2pt
                  \medskipamount=15pt plus4pt minus4pt
                  \bigskipamount=30pt plus8pt minus8pt
                  \normalbaselineskip=30pt plus0pt minus0pt
                  \normallineskip=2pt
                  \normallineskiplimit=0pt
                  \jot=7.5pt
                  {\def\smallskip {\vskip\smallskipamount}}
                  {\def\medskip   {\vskip\medskipamount}}
                  {\def\bigskip   {\vskip\bigskipamount}}
                  {\setbox\strutbox=\hbox{\vrule
                    height21.0pt depth9.0pt width 0pt}}
                  \parskip 15.0pt
                  \normalbaselines}

\tightenlines
\begin{document}
\preprint{
%\font\fortssbx=cmssbx10 scaled \magstep2
%\hbox to \hsize{
%\special{psfile=iulogo.ps
% hscale=8000 vscale=8000
% hoffset=-12 voffset=-2}
%\hskip.5in \raise.1in\hbox{\fortssbx University of Wisconsin - Madison}
\hfill$\vcenter{\hbox{\bf IUHET-407} \hbox{June
             1999}}$  }

\title{\vspace*{.75in}
Abelian Family Symmetries and Leptogenesis}

\author{M. S. Berger
\footnote{Electronic address:
berger@gluon.physics.indiana.edu}}

\address{
Physics Department, Indiana University, Bloomington, IN 47405, USA}

\maketitle

\thispagestyle{empty}

\begin{abstract}
We study the impact of a set of horizontal symmetries on the requirements
for producing the baryon asymmetry of the universe via leptogenesis.
We find that Abelian horizontal symmetries lead to a simple description of the
parameters describing leptogenesis in terms of the small expansion parameter
that arises from spontaneous symmetry breaking. If the family symmetry 
is made discrete, then an enhancement in the 
amount of leptogenesis can result. 
\end{abstract}

\newcommand{\be}{\begin{equation}}
\newcommand{\ee}{\end{equation}}
\newcommand{\bea}{\begin{eqnarray}}
\newcommand{\eea}{\end{eqnarray}}

\newpage

\section{Introduction}
There is now strong evidence for atmospheric neutrino oscillations.
The data suggests\cite{superk} that $\nu_\mu -\nu_\tau$ oscillations occur 
with near maximal 
mixing $\sin^2 2\theta _{23}\approx 1$ and a mass splitting of 
$\Delta m^2_{23}\sim 2.2\times 10^{-3}$~eV$^2$.
The measured solar neutrino flux can be explained by oscillations of $\nu _e$
to the other two generations ($x=2,3$). In the case of 
matter oscillations (MSW) there
are two solutions: (1) the small mixing angle (SMA) solution for which
$\Delta m^2_{1x}\sim 5\times 10^{-6}$~eV$^2$ and 
$\sin ^2 2\theta_{1x} \sim 6\times 10^{-3}$, and
(2) the large mixing angle (LMA) solution for which 
$\Delta m^2_{1x}\sim 2\times 10^{-5}$~eV$^2$ 
and $\sin ^2 2\theta_{1x} \sim 0.8$.
In the case of vacuum oscillations (VO) the mass-squared difference is much 
smaller $\Delta m^2_{1x}\sim 8\times 10^{-11}$~eV$^2$ 
and the mixing angle is also
large, $\sin ^2 2\theta_{1x} \sim 0.8$.
The largeness of the mixing $\theta _{23}$ and possibly in 
$\theta _{1x}$ and the apparent hierarchy 
in the associated masses presents something of a dilemma, since one would
expect that large mixing of order one occur when the eigenvalues (neutrino
masses) are roughly degenerate. Many models have been proposed 
to account for the neutrino oscillation data, and it is interesting to explore
whether these models can account in a natural way for the baryon asymmetry
of the universe through the process of leptogenesis. In this paper we
explore the implications for Abelian family symmetries on lepton asymmetries
generated in the early universe. In particular we argue that a discrete $Z_2$
component can not only resolve the dilemma of large mixing together 
with a large hierarchy mentioned above, but it can also lead to an 
enhanced baryon asymmetry.

\section{The Baryon Asymmetry and Leptogenesis}

The lightness of the three known neutrinos can be understood as arising from 
the see-saw mechanism where right-handed neutrinos, being Standard Model 
singlets, have a very large mass. The addition of right-handed singlet 
neutrinos to the Standard Model leads to lepton number violation.
The existence of very heavy right-handed neutrinos are predicted by
grand unified theories based on the gauge group $SO(10)$, and the lightness
of the observed neutrinos can be explained via a see-saw mechanism. 
Since the heavy right-handed neutrinos 
offer a reasonable basis for the observed oscillations and neutrino masses,
it motivates the consideration of their possible cosmological effects.
Since these particles would naturally occur in the early universe, it is of
interest to determine whether it is possible that the decays of these 
heavy particles could be the source of the baryon asymmetry of the 
universe\cite{fy}. 

The nonzero net baryon density $n_B-n_{\overline B}$ 
of the universe can be accounted for in theories that satisfy 
Sakharov's conditions\cite{Sakharov}: 1) baryon number is 
violated, 2) charge conjugation symmetry (C) and CP are violated, and 3) there
is a departure from thermal equilibrium. A nontrivial requirement on any 
particle theory satisfying these three conditions 
is that a sufficient asymmetry
in $n_B$ and $n_{\overline B}$ be produced 
to explain the observed value of the ratio of 
net baryon density to the entropy density 
$s$ of the universe 
\bea
&&Y_B={{n_B-n_{\overline B}}\over s}=(0.6-1)\times 10^{-10}\;. \label{asym}
\eea 
The Standard Model in the early universe satisfies all three conditions, but
it is generally agreed that the produced asymmetry is too small\cite{SM-bary}. 
Therefore one is motivated to look beyond 
the Standard Model at theories that contain
new sources of baryon number violation and CP-violation and/or for 
theories that
have a new mechanism for producing the asymmetry.
If one instead considers the Minimal Supersymmetric Standard Model (MSSM) then 
the regions of parameter space where sufficient baryon asymmetry is produced is
quite small\cite{MSSM-bary}.
Consequently various proposals have been made for new physics 
capable of  producing the baryon asymmetry of the 
universe. One of the most attractive of these is the possibility that 
CP violating decays of heavy neutrinos can produce an excess of leptons
over antileptons (or vice versa). The lepton asymmetry produced 
in the early universe via out-of-equilibrium 
decays of the right-handed neutrinos is subsequently recycled into 
a baryon asymmetry by sphaleron transitions (which violated both baryon
number and lepton number).    
A straightforward analysis of chemical potentials for equilibrating
processes including the sphaleron transition relates the baryon asymmetry 
$Y_B$ to the original lepton asymmetry 
$Y_L=(n_L-n_{\overline L})/s$ via\cite{ht,bp3}
\bea
&&Y_B=aY_{B-L}={a\over {a-1}}Y_L\;, 
\qquad a={{8N_F+4N_H}\over {22N_F+13N_H}}\;,
\eea
where $N_F$ is the number of fermion families and $N_H$ is the number of 
Higgs doublets. So the final baryon asymmetry present in the universe today
is related to the lepton asymmetry $Y_L$ by an order one parameter.
If one accepts the presence of heavy Majorana neutrinos in 
nature, then CP-violation naturally occurs and the question becomes whether
or not the lepton asymmetry that results is the right order of magnitude
for producing the observed baryon asymmetry in Eq.~(\ref{asym}). 
In the MSSM with heavy
right-handed neutrinos, 
the resulting lepton asymmetry has been shown to 
be sufficient to explain the observed baryon asymmetry in a natural way in 
a number of models\cite{bp2,plu-th,bp,by}.

Most work in trying to understand the structure of the fermion masses and 
mixings has tried to fit the low energy data, e.g. the fermion masses and 
the CKM matrix as well as the neutrino data (especially the solar neutrino 
oscillation data and the atmospheric neutrino oscillation data). If one
accepts the notion that leptogenesis is the source of the baryon asymmetry 
of the universe, then this mechanism imposes another rather strong constraint
on the details of the family symmetry (this kind of symmetry is also called
a horizontal symmetry). For example the lepton asymmetry 
produced by the decay of heavy Majorana neutrinos is sensitive to the 
texture pattern of the Yukawa matrices as well as the details of the mass and 
mixing hierarchies\cite{bb}. In the next section we apply the strategy 
of employing an Abelian family symmetry to describe the hierarchies and 
discuss the implications for leptogenesis.

\section{Horizontal Symmetries}
One attempt at accounting for the fermion mass spectrum makes use of broken 
family symmetries\cite{fn}. The most common approach is to take an Abelian 
$U(1)$ as the horizontal symmetry, but nonabelian groups and discrete groups
(and combinations of these) have been tried with varying degrees of success.
Since an Abelian symmetry alone cannot generate a nearly degenerate set of 
neutrinos\cite{bhkr}, we assume here that the 
$\Delta m_{23}^2$ and $\Delta m_{1x}^2$ are
indicating that the neutrino masses are arranged in a hierarchical pattern.
This hierarchical structure of the fermion masses suggests that it might be 
produced by an expansion in a small parameter, and one widely adopted 
strategy is to
have this parameter arise from a family symmetry spontaneously broken at a 
scale $\Lambda _L$. 
In this paper we consider the possibility that the horizontal 
symmetry
is an Abelian 
anomalous gauge symmetry\cite{br,nir}, where the anomaly is cancelled 
by the Green-Schwarz mechanism\cite{gs}. In this scenario there is field $\Phi$
that is a singlet under the Standard Model gauge symmetries. The contribution 
of the Fayet-Iliopoulos term to the D-term cancels against the contribution 
from the vev, $< \Phi >$. The ratio of this vev to the Planck scale naturally
provides a small parameter $\lambda = <S>/m_{\rm Pl}$. The field $\Phi$ is 
charged under the horizontal symmetry, and without loss of generality it 
charge can be taken to be $-1$. In this approach the hierarchy is generated by 
nonrenormalizable terms that transform as singlets under the horizontal 
symmetry and therefore produce contributions to the mass matrices that 
contain integer powers of the small parameter $\lambda$.
In this scenario it is often the case that only the $(3,3)$ entry of one or 
more mass 
matrices receives a contribution from a renormalizable coupling to the 
Higgs boson. So by assigning 
quantum numbers for the horizontal symmetry for each Standard Model field, one
can generate a hierarchy in the Yukawa matrices as powers of the small 
parameter $\lambda$.
 
The heavy Majorana neutrino mass matrix $M_N$ is obtained by inverting the
type-I see-saw formula 
\be
m_\nu=m_D (M_N)^{-1} m_D^T \label{eqn1}\;, \label{seesaw}
\ee
where $m_D$ is the neutrino Dirac mass matrix.
CP-asymmetries in neutrino decays arise from the interference
between the tree level and one-loop level decay channels. 
In the mass basis where the right-handed Majorana mass matrix is diagonal the 
asymmetry in heavy neutrino $N_i$ decays   
\bea
\epsilon_i&=&{{\Gamma (N_i\to \ell H_2)-\Gamma (N_i\to \ell ^c H_2^c)}\over
{\Gamma (N_i\to \ell H_2)+\Gamma (N_i\to \ell ^c H_2^c)}}\;,
\eea
is given by \cite{bp2,crv} 
\bea
\epsilon_i&=&{3\over {16\pi v_2^2}}{1\over {(m_D^\dagger m_D)_{ii}}}
\sum_{n\ne i}{\rm Im} \left [(m_D^\dagger m_D)_{ni}^2\right ]
{{M_i}\over {M_n}}\;.\label{epsilon}
\eea
The masses $M_i$ are the
three eigenvalues of the heavy Majorana mass matrix and $v_2$ is
the vev of the Higgs giving Dirac masses to the neutrinos and up-type 
quarks. 
$M_1$ is the
mass of the lightest of the three heavy Majorana neutrinos, and 
Eq.~(\ref{epsilon}) is an approximate formula 
valid for $M_n >> M_i$. The most common scenario that occurs 
is that the lightest Majorana neutrino $N_1$
has a mass such that $M_1<<M_2,M_3$,
and the lepton asymmetry produced\footnote{In some cases inverted 
hierarchies in the 
Majorana mass matrix 
can occur where $M_2<M_1$, which can produce a larger asymmetry if 
$\epsilon _2 >\epsilon _1$\cite{cfl}. We do not consider this possibility 
in this paper.} comes almost entirely from the 
decays of $N_1$. So the CP-asymmetry of most interest to the discussion of
lepton asymmetry generation is $\epsilon_1$.

The other
parameter of most interest is the mass parameter 
\bea
\tilde{m}_1&=&{{(m_D^\dagger m_D)_{11}}\over {M_1}}\;, \label{m1t}
\eea
which controls the decay width of the lightest right-handed neutrino $N_1$
since
\bea
&&\Gamma _{N_i}=\Gamma(N_i\to \ell H_2)+\Gamma (N_i\to \ell ^c H_2^c)
={1\over {8\pi}}(m_D^\dagger m_D)_{ii}{{M_i}\over {v_2^2}}\;,
\eea
and $\tilde{m}_1$ also largely controls the amount of dilution caused by the 
lepton number 
violating scattering. The parameter $\tilde{m}_1$ can therefore be called the 
dilution mass. These two constraints bound the possible values of
$\tilde{m}_1$ such that a sufficient asymmetry is produced to agree with 
Eq.~(\ref{asym}). The generated lepton asymmetry is given by 
\bea
&&Y_L={{n_L-n_{\overline L}}\over s}=\kappa {{\epsilon _1}\over g^*}\;,
\label{YL}
\eea
where $g^*$ is the number of light (effective) 
degrees of freedom in the theory 
($106~{3\over 4}$ in the Standard Model or $228~{3\over 4}$ in the MSSM), and 
$\kappa $ is a dilution 
factor that can be reliably calculated by solving the full Boltzmann 
equations. The dilution depends critically on the 
parameter $\tilde{m}_1$ because it governs the size of the most important 
Yukawa coupling in the $\Delta L=2$ scattering processes, as shown in 
Ref.~\cite{bp2}. 

\subsection{Leptogenesis with a $U(1)$ family symmetry}

Assume now that the lepton fields have charges under a $U(1)$ family symmetry
\bea
&&\begin{array}{c@{\quad}c@{\quad}c@{\quad}c@{\quad}c@{\quad}c
@{\quad}c@{\quad}c@{\quad}c}
e_{R1}^c & e_{R2}^c & e_{R3}^c & \ell_{L1} & \ell_{L2} & \ell_{L3} 
& \nu_{R1}^c & \nu_{R2}^c & \nu_{R3}^c \\
           E_1 & E_2 & E_3 & L_1 & L_2 & L_3 & {\cal N} _1 
                    & {\cal N} _2 & {\cal N} _3    \end{array}\;.\nonumber
\eea
We assume here that the quantum numbers satisfy the hierarchies
$E_1\geq E_2\geq E_3\geq 0$, $L_1\geq L_2\geq L_3\geq 0$, and 
${\cal N}_1\geq {\cal N}_2\geq {\cal N}_3\geq 0$ (This last condition will
guarantee that no 
light neutrino masses are enhanced because a 
right-handed neutrino mass is suppressed\cite{gns,lr,bhssw}).

Given lepton doublet charges $L_i$ and right-handed neutrino charges ${\cal N}_i$
one has the following pattern for the neutrino Dirac mass matrix 
\bea
&&m_D \sim \pmatrix{\lambda ^{L_1+{\cal N}_1} & \lambda ^{L_1+{\cal N}_2} 
                                             & \lambda ^{L_1+{\cal N}_3} \cr
                      \lambda ^{L_2+{\cal N}_1} & \lambda ^{L_2+{\cal N}_2} 
                                             & \lambda ^{L_2+{\cal N}_3} \cr
                      \lambda  ^{L_3+{\cal N}_1} & \lambda ^{L_3+{\cal N}_2} 
                                             & \lambda ^{L_3+{\cal N}_3}}v_2\;,
\eea
and the following pattern for the Majorana mass matrix
\bea
&&M_N \sim \pmatrix{\lambda ^{2{\cal N}_1} & \lambda ^{{\cal N}_1+{\cal N}_2} 
                                             & \lambda ^{{\cal N}_1+{\cal N}_3} \cr
                      \lambda ^{{\cal N}_1+{\cal N}_2} & \lambda ^{2{\cal N}_2} 
                                             & \lambda ^{{\cal N}_2+{\cal N}_3} \cr
                      \lambda  ^{{\cal N}_1+{\cal N}_3} & \lambda ^{{\cal N}_2+{\cal N}_3} 
                                             & \lambda ^{2{\cal N}_3}}\Lambda _L\;.
\eea
Then one obtains the following form for the light neutrino mass matrix via
the see-saw formula Eq.~(\ref{seesaw}) 
\bea
&&m_\nu \sim \pmatrix{\lambda ^{2L_1} & \lambda ^{L_1+L_2} 
                                             & \lambda ^{L_1+L_3} \cr
                      \lambda ^{L_1+L_2} & \lambda ^{2L_2} 
                                             & \lambda ^{L_2+L_3} \cr
                      \lambda  ^{L_1+L_3} & \lambda ^{L_2+L_3} 
                                             & \lambda ^{2L_3}}
{{v_2^2}\over {\Lambda_L}}\;, \label{lightnu}
\eea
Clearly if $L_2=L_3$ one can obtain ${\cal O}(1)$ mixing in the 2-3 
sector\cite{gn},
or if $L_2=-L_3$ one has a pseudo-Dirac neutrino and maximal mixing in
the 2-3 sector\cite{blpr}.\footnote{It is also possible that 
one has only an approximate
equality $L_2\approx \pm L_3$ in which case the mixing is not truly order one, 
but could be sufficiently large to be phenomenologically relevant without 
assuming accidental cancellations\cite{lr}.}

The Super-Kamiokande collaboration measurements of the atmospheric neutrino
flux indicates large mixing $\sin ^2 2\theta _{23}\sim 1$ and a mass-squared 
difference $\Delta m_{23}^2\sim 2\times 10^{-3}~{\rm eV}^2$. 
The SMA solution to the solar neutrino oscillations 
requires $\Delta m_{1x}^2\sim 5\times 10^{-6}~{\rm eV}^2$. 
If one assumes that the 
light neutrino masses are hierarchical, then one can identify
$m_{\nu_\tau}^2\sim 2\times 10^{-3}~{\rm eV}^2$
and $m_{\nu_\mu}^2\sim 5\times 10^{-6}~{\rm eV}^2$; 
it is then difficult to naturally
explain the separation of masses simultaneously with the large mixing angle.
The suppression of one of the neutrino masses can always result from a 
fine-tuning of the parameters.

The dilution parameter $\tilde{m}_1$ defined in Eq.~(\ref{m1t})
can be described in terms of the $U(1)$ quantum numbers by constructing 
the Yukawa coupling squared matrix
\bea
&&m_D^\dagger m_D \sim \pmatrix{\lambda ^{2{\cal N}_1} & \lambda ^{{\cal N}_1+{\cal N}_2} 
                                             & \lambda ^{{\cal N}_1+{\cal N}_3} \cr
                      \lambda ^{{\cal N}_1+{\cal N}_2} & \lambda ^{2{\cal N}_2} 
                                             & \lambda ^{{\cal N}_2+{\cal N}_3} \cr
                      \lambda  ^{{\cal N}_1+{\cal N}_3} & \lambda ^{{\cal N}_2+{\cal N}_3} 
                                             & \lambda ^{2{\cal N}_3}}
\lambda ^{2L_3}v_2^2\;,
\eea
so that
\bea
&&\tilde{m}_1\sim {{\lambda ^{2(L_3+{\cal N}_1)}v_2^2}\over {M_1}}
\sim {{\lambda ^{2(L_3+{\cal N}_1)}}\over {\lambda ^{2{\cal N}_1}}}
{{v_2^2}\over {\Lambda _L}}\sim \lambda ^{2L_3}{{v_2^2}\over {\Lambda _L}}\;,
\label{dilutem}
\eea
When $L_2=L_3$ then this parameter is the same order of magnitude as the 
neutrino masses $m_{\nu_\mu}$ and $m_{\nu_\tau}$\footnote{We use the notation
$\nu_\mu$ and $\nu_\tau$ for the eigenstates even though they have large 
mixing.}, and 
it is consistent to take the parameter 
$\tilde{m}_1\sim (m_{\nu_\mu}m_{\nu_\tau})^{1/2}$. Typically one
needs a fine-tuning to produce the hierarchy $m_{\nu_\mu}<<m_{\nu_\tau}$.
The CP-violating parameter is given by 
\bea
&&\epsilon_1\sim {3\over {16\pi}}\lambda ^{2(L_3+{\cal N}_1)}\;.
\label{epsilon1}
\eea
Comparing to Eq.~(\ref{dilutem}), one sees that $\epsilon_1$ can be simply 
expressed in terms of the dilution mass $\tilde{m}_1$, the mass $M_1$ of the
lightest Majorana neutrino, and the electroweak scale vev $v_2$. Since 
$\tilde{m}_1$ is tied to the light neutrino masses, a connection between
these quantities is established at the order-of-magnitude level.

The problem with the situation outlined is well-known: it seems to predict 
that $m_{\nu_\mu}$ is the naturally of the same order as $m_{\nu_\tau}$, 
and one would need to have an accidental cancellation to get the hierarchy 
$m_{\nu_\mu}<<m_{\nu_\tau}$. The charged lepton matrix is given by 
\bea
&&m_{\ell^\pm} \sim \pmatrix{\lambda ^{L_1+E_1} & \lambda ^{L_1+E_2} 
                                             & \lambda ^{L_1+E_3} \cr
                      \lambda ^{L_2+E_1} & \lambda ^{L_2+E_2} 
                                             & \lambda ^{L_2+E_3} \cr
                      \lambda  ^{L_3+E_1} & \lambda ^{L_3+E_2} 
                                             & \lambda ^{L_3+E_3}}v_1\;,
\eea
where $v_1$ is the vev of the other Higgs doublet.
So the relevant rotation to get to the basis where the charged lepton mass 
is diagonal is also order one when $L_2=L_3$. Hence the large mixing in the
2-3 sector is connected in this approach to near degeneracy of two of the 
light neutrino masses. 

\subsection{Leptogenesis with a $Z_2\times U(1)$ family symmetry}

Ref.~\cite{gns} proposed that a discrete Abelian family symmetry could be
employed to enhance a mass or mixing angle above what would be otherwise 
obtained if the family symmetry was the usual continuous $U(1)$ symmetry, and
this idea was pursued further in a specific model\cite{tan}.
If the family symmetry is $Z_m$ then entries in the mass matrices can 
be enhanced by factors of the small parameter $\lambda$ to the $m$th power.
With this approach 
the discrete $Z_m$ symmetry can result in the enhancement of 
entries in the light neutrino mass matrix. A consequence
for leptogenesis is that this will also change 
the relationship between the light neutrino masses and the dilution parameter
$\tilde{m}_1$ by a factor of $\lambda ^m$.
For example take the following $Z_2\times U(1)$ charges for the lepton 
fields\footnote{The second group factor does not need to be continuous, but 
could be replaced by a second $Z_n$ with $n$ sufficiently large.}
\bea
&&\begin{array}{c@{\quad}c@{\quad}c@{\quad}c@{\quad}c
@{\quad}c@{\quad}c@{\quad}c@{\quad}c}
e_{R1}^c & e_{R2}^c & e_{R3}^c & \ell_{L1} & \ell_{L2} & \ell_{L3} 
& \nu_{R1}^c & \nu_{R2}^c & \nu_{R3}^c \\
           (0,E_1) & (0,E_2) & (0,E_3) & (0,L_1) & (0,L_2) & (1,L_3-1) 
& (0,{\cal N}_1) & (0,{\cal N}_2) & (1,{\cal N}_3-1)    \end{array}\nonumber
\eea
and later we will take $L_2=L_3$. Assume the symmetry breaking is 
characterized by the single expansion parameter $\lambda$.
The formulas given above for the heavy neutrino mass matrix, $M_N$, 
the neutrino Dirac mass matrix, $m_D$, and the resulting light neutrino mass 
matrix, $m_\nu$ are modified. With the above assignments one finds that 
\bea
&&M_N \sim \pmatrix{\lambda ^{2{\cal N}_1} & \lambda ^{{\cal N}_1+{\cal N}_2} 
                                             & \lambda ^{{\cal N}_1+{\cal N}_3} \cr
                      \lambda ^{{\cal N}_1+{\cal N}_2} & \lambda ^{2{\cal N}_2} 
                                             & \lambda ^{{\cal N}_2+{\cal N}_3} \cr
                      \lambda  ^{{\cal N}_1+{\cal N}_3} & \lambda ^{{\cal N}_2+{\cal N}_3} 
                                             & \lambda ^{2{\cal N}_3-2}}\Lambda _L\;,
\eea
so that 
\bea
&&(M_N)^{-1} \sim \pmatrix{\lambda ^{-2{\cal N}_1} & \lambda ^{-{\cal N}_1-{\cal N}_2} 
                                             & \lambda ^{-{\cal N}_1-{\cal N}_3+2} \cr
                      \lambda ^{-{\cal N}_1-{\cal N}_2} & \lambda ^{-2{\cal N}_2} 
                                             & \lambda ^{-{\cal N}_2-{\cal N}_3+2} \cr
                      \lambda  ^{-{\cal N}_1-{\cal N}_3+2} & \lambda ^{-{\cal N}_2-{\cal N}_3+2} 
                                             & \lambda ^{-2{\cal N}_3+2}}\Lambda _L^{-1}\;.
\eea
Furthermore one has
\bea
&&m_D \sim \pmatrix{\lambda ^{L_1+{\cal N}_1} & \lambda ^{L_1+{\cal N}_2} 
                                             & \lambda ^{L_1+{\cal N}_3} \cr
                      \lambda ^{L_2+{\cal N}_1} & \lambda ^{L_2+{\cal N}_2} 
                                             & \lambda ^{L_2+{\cal N}_3} \cr
                      \lambda  ^{L_3+{\cal N}_1} & \lambda ^{L_3+{\cal N}_2} 
                                             & \lambda ^{L_3+{\cal N}_3-2}}v_2\;,
\eea
Then it is straightforward to show that the light neutrino mass matrix $m_\nu$
is modified so that only one component is enhanced,
\bea
&&m_\nu \sim \pmatrix{\lambda ^{2L_1} & \lambda ^{L_1+L_2} 
                                             & \lambda ^{L_1+L_3} \cr
                      \lambda ^{L_1+L_2} & \lambda ^{2L_2} 
                                             & \lambda ^{L_2+L_3} \cr
                      \lambda  ^{L_1+L_3} & \lambda ^{L_2+L_3} 
                                             & \lambda ^{2L_3-2}}
{{v_2^2}\over {\Lambda_L}}\;. \label{lightnuenhance}
\eea
Also one finds that
\bea
&&m_D^\dagger m_D \sim \pmatrix{\lambda ^{2{\cal N}_1} & \lambda ^{{\cal N}_1+{\cal N}_2} 
                                             & \lambda ^{{\cal N}_1+{\cal N}_3-2} \cr
                      \lambda ^{{\cal N}_1+{\cal N}_2} & \lambda ^{2{\cal N}_2} 
                                             & \lambda ^{{\cal N}_2+{\cal N}_3-2} \cr
                      \lambda  ^{{\cal N}_1+{\cal N}_3-2} & \lambda ^{{\cal N}_2+{\cal N}_3-2} 
                                             & \lambda ^{2{\cal N}_3-4}}
\lambda ^{2L_3}v_2^2\;.\label{mddagmd_z2}
\eea
So then using our previous definitions, one sees that 
$\tilde{m}_1\sim \lambda ^{2L_3}v_2^2/\Lambda _L\sim m_{\nu_\mu}$, whereas 
$m_{\nu_\tau}\sim \lambda ^{2L_3-2}v_2^2/\Lambda _L$. More specifically when
the atmospheric neutrino constraint 
$\Delta m_{23}^2\sim 2\times 10^{-3}$~eV$^2$ is interpreted as the 
mass-squared of the heaviest light neutrino $m_{\nu_\tau}$, then in the case 
of a horizontal $U(1)$ symmetry, one has that 
$\tilde{m}_1^2\sim 2\times 10^{-3}$~eV$^2$. 
In the case of the discrete $Z_2$ symmetry,
the Yukawa coupling related to $\tilde{m}_1^2$ via Eq.~(\ref{m1t}) can be 
reduced by a factor $\lambda ^2$ thereby substantially reducing the amount
of dilution from the $\Delta L=2$ processes and reducing the decay rate of 
$N_1$. More generally, a $Z_m$ symmetry
can arrange for a suppression of $\tilde{m}_1^2$ by a factor $\lambda ^m$.
The CP-violation asymmetry $\epsilon_1$ is easily obtained from 
Eqs.~(\ref{epsilon}) and (\ref{mddagmd_z2}),
\bea
&&\epsilon_1\sim {3\over {16\pi}}\lambda ^{2(L_3+{\cal N}_1)-2}\;.
\label{epsilon1_z2}
\eea
So the ratio of $m_{\nu_\tau}$ to $\epsilon_1$ is unaffected by the discrete
symmetry. Since $m_{\nu_\tau}$ is being fixed by the experimental data
for $\Delta m_{23}^2$, the expected value of $\epsilon _1$ is expected to 
be unchanged for the same quantum number ${\cal N}_1$ (and thus the same
heavy neutrino mass $M_1$).

A phenomenologically viable solution has been presented in
Ref.~\cite{gns}: taking $L_1=3$, $L_2=L_3=1$, $E_1=5$, $E_2=4$, and $E_3=2$
yields mass matrices of the form
\bea
&&m_\nu \sim \pmatrix{\lambda ^6 & \lambda ^4 
                                             & \lambda ^4 \cr
                      \lambda ^4 & \lambda ^2 
                                             & \lambda ^2 \cr
                      \lambda  ^4 & \lambda ^2 
                                             & 1}
{{v_2^2}\over {\Lambda_L}}\;, 
\qquad
m_{\ell^\pm} \sim \pmatrix{\lambda ^8 & \lambda ^7 
                                             & \lambda ^5 \cr
                      \lambda ^6 & \lambda ^5 
                                             & \lambda ^3 \cr
                      \lambda  ^6 & \lambda ^5 
                                             & \lambda ^3}v_1\;,
\eea
which give the correct order of magnitude for the SMA 
solution (after rotating to the charged lepton mass basis)
\bea
{{\Delta m^2_{1x}\over\Delta m^2_{23}}\sim\lambda^4,
\ \ \ \sin \theta_{12}\sim\lambda^2,
\ \ \ \sin \theta_{23}\sim1,\ \ \ \sin \theta_{13}\sim\lambda^2,}
\eea
when the small parameter is identified as the Cabibbo angle, i.e. 
$\lambda \sim 0.2$.

For a sufficient amount of leptogenesis to occur two conditions must be 
satisfied: (1) $|\epsilon_1| \agt 10^{-6}$ and 
(2) $10^{-5} \alt \tilde{m}_1 \alt 10^{-2}$.
The first condition guarantees that there is sufficient CP-violation in the
heavy $N_1$ neutrino decay (c.f. Eq.~(\ref{YL}), 
while the second condition guarantees that the 
dilution is not too large ($\kappa \agt 10^{-2}$) and that a sufficient 
number of heavy neutrino are produced out-of-equilibrium\cite{plu-th}.
The condition on $\tilde{m}_1$ is equivalent to a condition on the relevant
Yukawa coupling $(h_\nu ^\dagger h_\nu)_{11}=m_D^\dagger m_D/v_2^2$
that governs the rates of these two processes.

The mass $\tilde{m}_1$ 
arising in the case of the $U(1)$ symmetry is identified with $m_{\nu_\tau}$, 
and thus is near the top of the required
range. The resulting lepton asymmetry is smaller than it would be 
if the dilution mass $\tilde{m}_1$ could be reduced.
Lowering the mass parameter $\tilde{m}_1$ by using the horizontal 
$Z_2\times U(1)$ rather
than the $U(1)$ symmetry has the following effects on 
the Boltzmann evolution: 1) The lightest Majorana neutrino $N_1$
decays more slowly, and stays out of thermal equilibrium for a longer period
of time. 2) The dilution of the generated lepton asymmetry is reduced since 
the relevant Yukawa coupling controlling the strength of the interactions 
is reduced. These two factors can result in a remnant lepton asymmetry that
is enhanced over that which is obtained in the case of the $U(1)$ symmetry.

\section{Numerical Simulation}

The lepton asymmetry that results can be obtained by integrating the 
full set of Boltzmann equations\cite{fot}. These differential
equations, incorporating the Majorana neutrino decay rates as well as all 
lepton number violating scattering processes in the MSSM, has been given in 
Ref.~\cite{plu-th}. The above discussion gives an overall order of magnitude
estimate for the CP-violation parameter $\epsilon _1$ and the dilution 
parameter $\tilde{m}_1$. The parameter $\epsilon _1$ depends on a
CP-phase (see Eq.~(\ref{epsilon})); this phase is not specified by the family
symmetry and we assume that it is order one.

A concrete
example of how the discrete symmetry can change the produced lepton asymmetry 
is shown in Figs.~1 and 2. First consider the case where the family symmetry
is $U(1)$: 
the values of the parameters are 
$(m_D^\dagger m_D)_{11}=0.2$~GeV$^2$
and $\epsilon_1=-4.0 \times 10^{-6}$ for the case of the continuous $U(1)$
symmetry. The values for 
the CP-violation parameter $\epsilon_1$ in Eq.~(\ref{epsilon1}) and 
$(m_D^\dagger m_D)_{11}\sim \lambda ^{2(L_3+{\cal N}_1)}v_2^2$ are
consistent\footnote{These 
relationships involving $\epsilon _1$ and $(m_D^\dagger m_D)_{11}$
only determine the leading
order contribution in the small parameter $\lambda$ and there is an 
undetermined coefficient of order one. We choose the values of the parameters
here as an example to illustrate that an enhancement occurs. The exact values
the unknown order one coefficients take are not important; the enhancement of 
the lepton asymmetry is generic.}
with taking 
$L_3=0$ and ${\cal N}_1=3$.
Taking the scale $\Lambda _L$ to be near a supersymmetric grand unified scale
$\sim 3\times 10^{15}$~GeV, one finds the mass of the lightest heavy neutrino
$N_1$ that is decaying asymmetrically to be $M_1=2\times 10^{11}$~GeV.
This then yields a 
dilution mass of
$\tilde{m}_1=5\times 10^{-3}$~eV,
which is the same order of magnitude as the mass splitting 
$\Delta m_{23}^2\sim 2.2\times 10^{-3}$~eV$^2$
as expected from Eqs.~(\ref{lightnu}) and (\ref{dilutem}). 

\begin{center}
\epsfxsize=5.5in
\hspace*{0in}
\epsffile{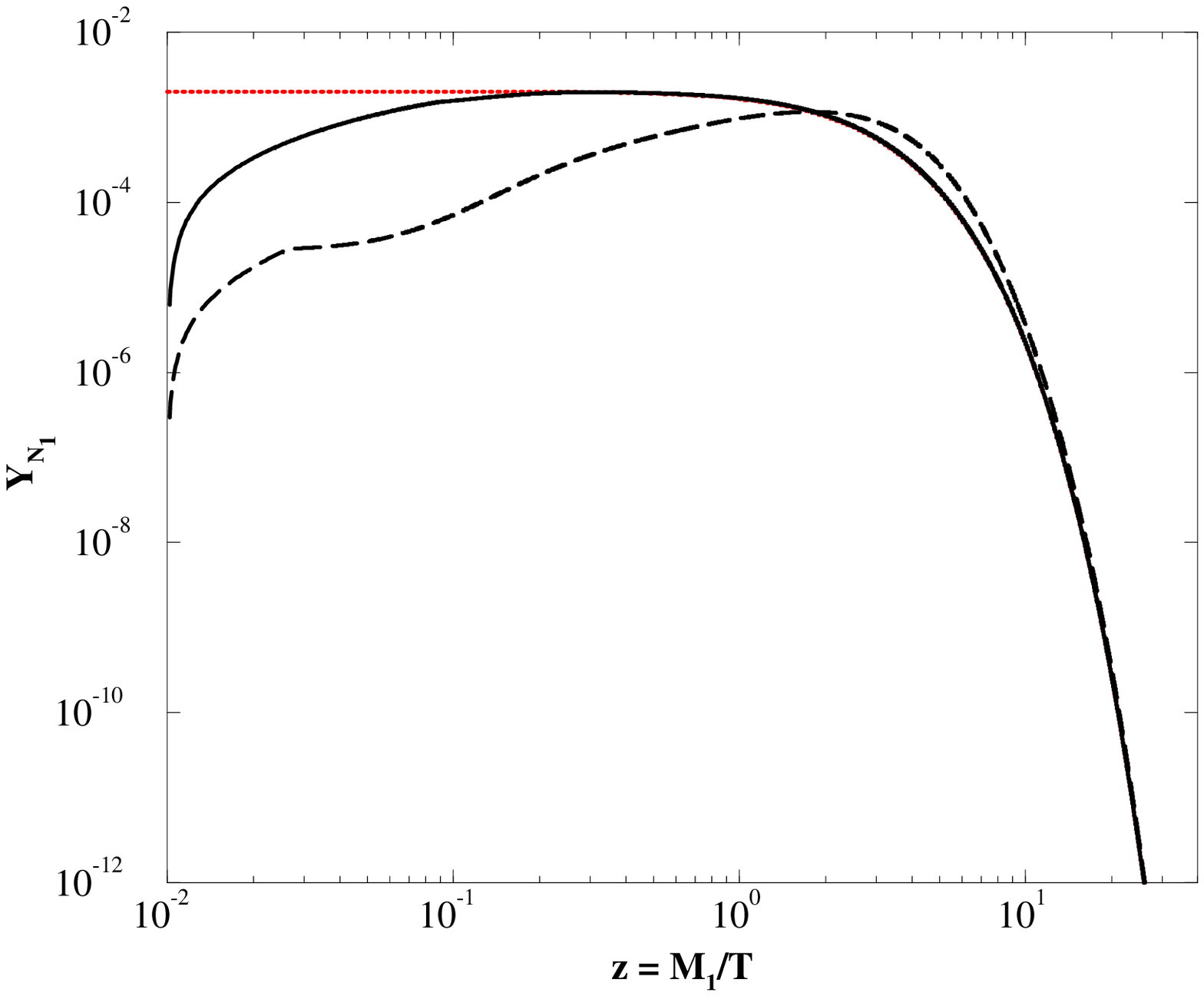}
\vspace*{0in}

\smallskip
\parbox{5.5in}{\small  Fig.~1. The neutrino density $Y_{N_1}$ as a function
of the temperature $T$ of the universe for the case of 
a horizontal $U(1)$ symmetry (solid), and for the $Z_2\times U(1)$ 
symmetry (dashed). The dotted curve is the equilibrium value $Y_{N_1}^{\rm eq}$
of the neutrino density. The discrete symmetry results in a smaller decay
rate for $N_1$ and it requires a longer time before it comes into thermal 
equilibrium. }
\end{center}
 
When the $U(1)$ 
symmetry is replaced with $Z_2$ 
the dilution mass is suppressed by an additional 
factor of $\lambda ^2$ so that $\tilde{m}_1=2.2\times 10^{-4}$~eV (for
the case of the $Z_2\times U(1)$ symmetry, we take $L_3=1$ and keep 
${\cal N}_1=3$ so that $m_{\nu_\tau}$ and $\epsilon_1$ remain the same, but
$\tilde{m}_1$ is reduced by a factor $\lambda ^2$ relative to the $U(1)$ 
symmetry case.).
Figure 1 shows the neutrino density $Y_{N_1}$ of 
the lightest Majorana neutrino that is decaying to produce to produce the 
lepton asymmetry shown in Fig.~2. The densities are plotted against the 
dimensionless ratio $z=M_1/T$ where $T$ is the temperature of the universe, 
so the universe evolves toward the present day as $z$ becomes larger.
For the quantitative results shown in the figures, the unknown CP phase 
(see Eq.~(\ref{epsilon}) is chosen so as to maximize the lepton asymmetry;
another phase would just scale the curves in Fig.~2 by some overall factor. 
The $Z_2$ symmetry results in $N_1$ decaying
more slowly, and thus $N_1$ can remain out-of-equilibrium for a greater 
period of time in the early universe. The lepton asymmetry produced in each
case begins with one sign, then goes through zero, and finally asymptotes to 
a final value. For the particular example shown in Figs. 1 and 2, the
lepton asymmetry is enhanced by about a factor between seven and eight
when the continuous family
symmetry is replaced by a discrete one. The enhancement (or suppression) that
can result in general (from suppressing $\tilde{m}_1$) 
is a sensitive function of the values of dilution 
parameter $\tilde{m}_1$ and the mass $M_1$, as shown in Ref.~\cite{plu-th}. 

\begin{center}
\epsfxsize=5.5in
\hspace*{0in}
\epsffile{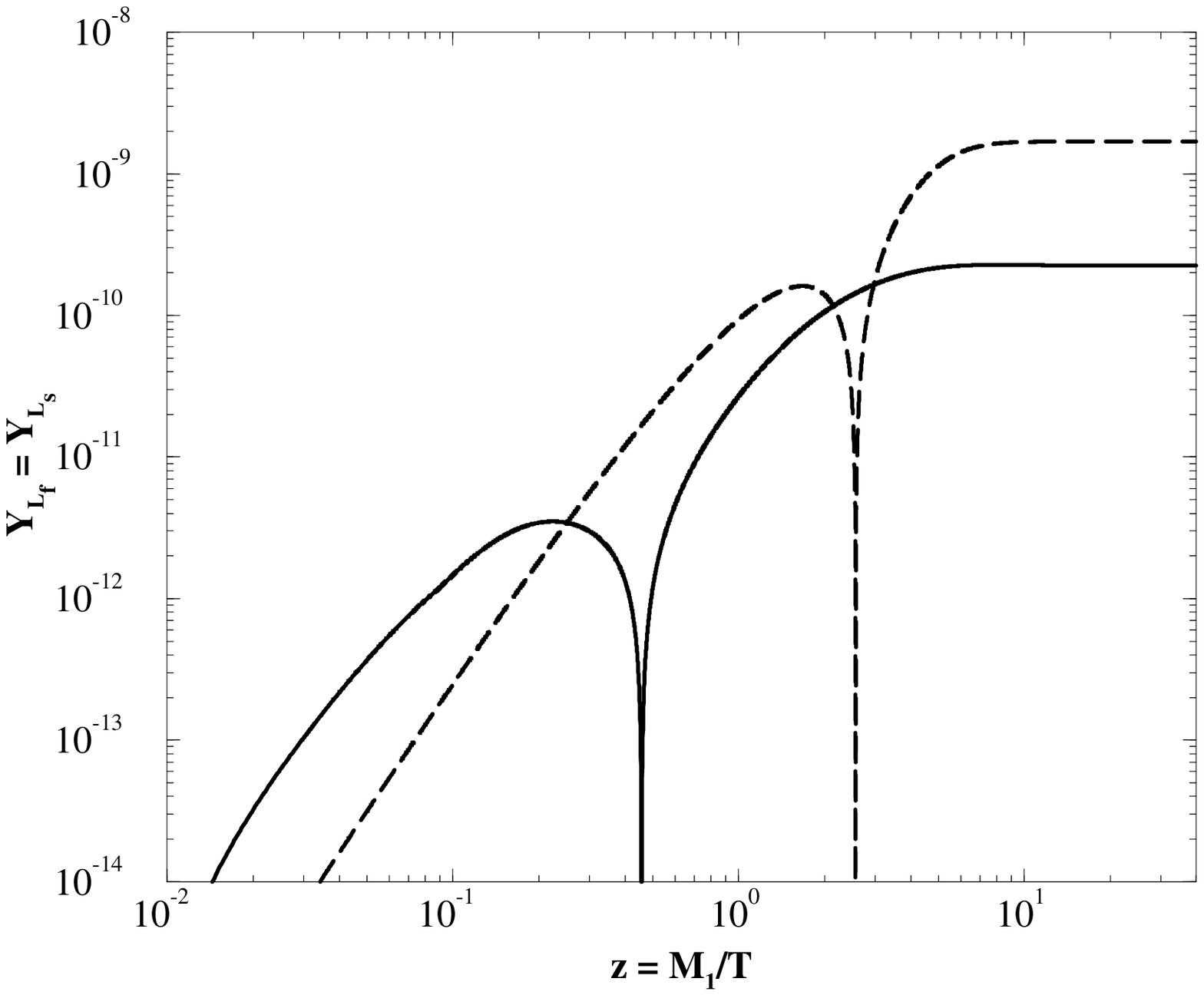}
\vspace*{0in}

\smallskip
\parbox{5.5in}{\small  Fig.~2. The lepton asymmetry in fermions $Y_{L_f}$ 
and in scalars $Y_{L_s}$ produced
for a horizontal $U(1)$ symmetry (solid), and for the $Z_2\times U(1)$ 
symmetry (dashed). The generated asymmetry in the latter case is smaller at 
earlier times (larger temperatures) since the decay rate of the lightest
Majorana neutrino $N_1$ is suppressed, but ultimately a larger asymmetry is
produced as the neutrino density remains out of thermal equilibrium for a 
longer period. The equality $Y_{L_f}=Y_{L_s}$ is maintained by MSSM processes
$f+f\leftrightarrow \tilde{f}+\tilde{f}$,
e.g. neutralino exchange.}
\end{center}
 
\section{Conclusion}
The predominance of matter over antimatter in the universe can be produced 
CP-violation in the decays of heavy neutrinos followed by sphaleron processes
that recycle the resulting lepton asymmetry into a baryon asymmetry.
We have shown that if the fermion mass matrices are determined by 
imposing an Abelian 
family symmetry then there are simple order-of-magnitude estimates of the 
CP-violation parameter $\epsilon_1$ and the dilution mass $\tilde{m}_1$ that
are critically important for determining the size of the lepton asymmetry
produced in the early universe. In the most straightforward case these 
parameters are given by universal formulas in terms of the $U(1)$ quantum 
numbers ($\epsilon _1\sim (3/16\pi)\lambda ^{2({\cal N}_1+L_3)}$ and 
$\tilde{m}_1\sim \lambda ^{2L_3}v_2^2/\Lambda_L$), and $\tilde{m}_1$ 
can be simply related to the experimentally determined light neutrino masses.

A $Z_2$ horizontal symmetry can be employed to reconcile (a) the large mixing
that must be present to explain the atmospheric neutrino data with (b) a 
hierarchy in neutrino masses. We have shown here that employing this 
same $Z_2$ horizontal symmetry can enhance the 
lepton asymmetry that results from heavy right-handed neutrino decays.
This results in an enhanced baryon asymmetry in the universe.
The change in the generated lepton 
asymmetry comes about because when a Yukawa coupling 
$(h_\nu ^\dagger h_\nu)_{11}$ can be suppressed or enhanced compared to the
usual expectation when the horizontal symmetry is $U(1)$. This affects the 
decay rate of the lightest Majorana neutrino $N_1$, as well as the amount of
subsequent dilution of the asymmetry by lepton number violating scattering.
A particular example where the generated 
asymmetry was explicitly calculated using the supersymmetric Boltzmann 
equations was given, and an enhancement of the lepton asymmetry (and hence 
ultimately the baryon asymmetry) by a factor seven
was derived quantitatively.

\section*{Acknowledgments}

This work was supported in part by the U.S.
Department of Energy
under Grant No. 
No.~DE-FG02-91ER40661.

\end{document}